\documentclass[11pt]{article}
\usepackage{amssymb}
\usepackage{epsfig}
\begin{document}
\title{{\bf{\Large Vacuum condition and the relation between response parameter and anomaly coefficient in ($1+3$) dimensions}}}
\author{ 
 {\bf {\normalsize Bibhas Ranjan Majhi}$
$\thanks{E-mail: bibhas.majhi@mail.huji.ac.il}}\\ 
{\normalsize Racah Institute of Physics, Hebrew University of Jerusalem,}
\\{\normalsize Givat Ram, Jerusalem 91904, Israel}
\\[0.3cm]
}
\maketitle

\begin{abstract}
The role of Israel-Hartle-Hawking vacuum is discussed for anomalous fluid in presence of both the gauge and gravitational anomalies in ($1+3$) dimensions. I show that imposition of this vacuum condition leads to the relation $\tilde{c}_{4d}=-8\pi^2c_m$ between the response parameter ($\tilde{c}_{4d}$) and the anomaly coefficient ($c_m$). This establishes a connection between the coefficients appearing in a first order and a third order derivative terms in the constitutive relation. 
\end{abstract}

\section{Introduction}
   Anomalies are the intrinsic properties of quantum field theory and they play an important role in various branches of physical phenomenon. Recently, it has been observed that the hydrodynamics, in presence of anomalies (gravitation or gauge or both), modifies nontrivially \cite{Rangamani:2009xk} -- \cite{Jimenez-Alba:2014pea}. Normally the hydrodynamics of a system is governed by certain constitutive relations, like energy-momentum tensor, current expressed in terms of the fluid variables. These are constructed in such a way that the theory comes out to be compatible with the local version of the second law of thermodynamics, which tells that the production of entropy must be positive. Usually they are found by derivative expansion method. We showed that in ($1+1$) dimensions, the constitutive relations can be obtained exactly in presence of both gravitational and gauge anomalies where perturbative approach is not needed \cite{Banerjee:2013qha,Banerjee:2013fqa}. The chiral theory leads to the stress-tensor which is identical to the ideal (chiral) fluid in form. Of course, these results agree with the derivative expansion method in absence of gauge fields.

   One of the important observations in the context of anomalous fluid tells that the anomalies contribute to the response parameters at two orders of derivatives of fluid variables less in the constitutive relations \cite{Jensen:2012kj}. It has been argued that such feature is very general and happens in any spacetime dimensions \cite{Loganayagam:2012pz}. For instance, in ($1+1$) dimensions the response parameters, enter at zeroth order, are proportional to the anomaly coefficients appearing in the terms which are second order derivatives of fluid variables \cite{Jensen:2012kj,Majhi:2014eta}. Similarly, in ($1+3$) dimensional case the same occurs among the first order and the third order terms \cite{Jensen:2012kj,Jensen:2013rga}. It turns out that these constraints are geometric rather than algebraic and the proportionality constant is quite universal, independent of spacetime dimensions.

  Recently, in establishing these universal relations the role of the vacuum condition for the quantum field theory in curved spacetimes has been illuminated by me \cite{Majhi:2014eta}. I showed that for ($1+1$) dimensional fluid, in presence of both diffeomorphism and trace anomalies (gravitational), the {\it Israel-Hartle-Hawking vacuum} is the relevant boundary condition to obtain this general feature. The intriguing fact of the analysis is that it enlightens the importance of vacuum conditions in hydrodynamics paradigm, like the role played by the Unruh vacuum in the context of Hawking radiation \cite{Banerjee:2008wq, Banerjee:2008sn}. This was not emphasised in the earlier discussions. Moreover, the whole computation is technically simple. My idea has also been followed to find the corrections to these constraints in presence of $U(1)$ gauge anomaly in two dimensions \cite{Banerjee:2014cya}. The important outcome is that one obtains the explicit expression for the gauge contribution which was failed to determine in the earlier analysis.

  It may be emphasised that to have better understandings on the roles of vacuums, one needs to look for the validity of the method in higher dimensions. The present manuscript precisely addresses this issue. In this short paper, I will extend my analysis in ($1+3$) dimensional case in presence of mixed anomalies; i.e. the theory has a mixture of gravitational and $U(1)$ gauge contributions in the anomaly equations. Unfortunately, four dimensional equations can not be solved exactly to find the constitutive relations (i.e. stress-tensor and current) like those in the two spacetime dimensions \cite{Banerjee:2013qha,Banerjee:2013fqa,Majhi:2014eta}. So one needs the perturbative approach. This has been precisely done in \cite{Jensen:2012kj,Jensen:2013rga} by derivative expansion method. The expressions are available up to third order. Here I shall borrow them and show that imposition of the {\it Israel-Hartle-Hawking vacuum} on the components of energy-momentum tensor in null coordinates leads to identical connection:
\begin{equation}
\tilde{c}_{4d} = -8\pi^2c_m~,
\label{main}
\end{equation}
where $\tilde{c}_{4d}$ is the response parameter while $c_m$ is the anomaly coefficient. In the below, I shall present the main analysis to derive the above result.

   The organization of the paper is as follows. In section \ref{summary} the expressions for the anomalous constitutive relations in four dimensions will be given up to third order in derivatives of the fluid variables. Next section will discuss the derivation of the required relation by using the relevant vacuum condition. Finally, I shall conclude in section \ref{conclusions}.

\section{\label{summary}Constitutive relations: summary of the results}
   In this section, I shall give a brief summary of the expressions for the stress-tensor and the current in presence of mixed anomalies without the details. These will be used in the next section to obtain the main result of the paper. 

   In ($1+3$) dimensions the covariant form of the anomaly equations, in presence of both the gravity and $U(1)$ gauge fields, are \cite{Jensen:2012kj},
\begin{eqnarray}
&&\nabla_aJ^a = \frac{1}{4}\epsilon^{abcd}\Big[3c_A F_{ab}F_{cd}+c_m R^i_{~~jab} R^j_{~~icd}\Big]~;
\label{1.01}
\\
&& \nabla_b T^{ab} = F^a_{~~b}J^b + 2c_m\nabla_b\Big[\frac{1}{4}\epsilon^{ijmn}F_{ij}R^{ab}_{~~~mn}\Big]~,
\label{1.02}
\end{eqnarray}
where $c_A$ is the $U(1)^3$ triangle anomaly coefficient while $c_m$ is the mixed (i.e. $U(1)$ and gravitational) anomaly coefficient. 
Note that gravitational and gauge anomalies are mixed up and they are distributed symmetrically in the anomaly for current. The above equations can not be solved exactly to find the corresponding constitutive relations for the current ($J^a$) and energy-momentum tensor ($T^{ab}$) like in ($1+1$) dimensional theory. Of course, it is possible to find them up to some orders in derivative of the fluid variables by the derivative expansion method. In literature, the results up to third order derivative in fluid variables are available \cite{Jensen:2012kj}. In absence of the exact expressions we will restrict our discussions within the third order derivative. The current and the stress-tensor can be found out by varying the generating function. This has been done in \cite{Jensen:2012kj}. Let me just summarize the main results below. The expressions of these quantities are \cite{Jensen:2012kj}
\begin{eqnarray}
&&J^{a} = J^{a}_{(0)} + J^{a}_{(A)}~;
\label{1.03}
\\
&&T^{ab} = T^{ab}_{(0)}+T^{ab}_{(A)}~;
\label{1.04} 
\end{eqnarray}
where
\begin{eqnarray}
&&J^{a}_{(0)} = \mathcal{N}u^a+\nu^{a}~;
\label{1.05}
\\
&& T^{ab}_{(0)} = \mathcal{E}u^au^b + \mathcal{P}\Delta^{ab}+u^aq^b+u^bq^a+\tau^{ab}~; 
\label{1.06}
\\
&& J^a_{(A)} = \mathcal{N}_{A}u^a+\nu^{a}_A~;
\label{1.07}
\\
&& T^{ab}_{(A)} = \mathcal{E}_Au^au^b + \mathcal{P}_A\Delta^{ab}+u^aq^b_A+u^bq^a_A+\tau^{ab}_A~.
\label{1.08} 
\end{eqnarray}
In the above $u^a$ is the fluid velocity satisfies the normalization $u_au^a=-1$ and the individual expressions for each term on the right hand side are as follows:
\begin{eqnarray}
&&\mathcal{P} = P-\zeta\nabla_au^a; \,\,\,\ \mathcal{E} = -P+\mu\frac{\partial P}{\partial\mu} + T\frac{\partial P}{\partial T}; \,\,\ 
\mathcal{N} = \frac{\partial P}{\partial\mu}~;
\nonumber
\\
&& \nu^a = \tilde{c}_{4d}T^2 \omega^a + \sigma\Delta^{ab}(E_b - T\nabla_b\frac{\mu}{T})~;
\nonumber
\\
&& q^a = \tilde{c}_{4d}T^2 B^a + 2\tilde{c}_{4d}\mu T^2 \omega^a~; 
\nonumber
\\
&& \tau^{ab} = -\eta\sigma^{ab}~,
\label{1.09}
\end{eqnarray}
and
\begin{eqnarray}
&& \mathcal{N}_A = 0; \,\,\,\ \mathcal{E}_A = 3\mathcal{P}_A = 2c_m\omega_a(\mu v^a_{2,1}-v^a_{2,2})+4c_mB_a(v^a_{2,1}-\Omega^{ab}a_b)~;
\nonumber
\\
&& \nu^a_A = -6c_A\mu B^a - 3c_A\mu^2\omega^a - 4c_mt^{ab}_2\omega_b - \frac{4}{3}c_m\Big(s_{2,1}-\frac{9}{2}\omega^2\Big)\omega^a~;
\nonumber
\\
&& q^a_A = -3c_A\mu^2B^a-2c_A\mu^3\omega^a-2c_m\Big(\Phi^{ab}B_b+(\omega^2/4-a^2)B^a\Big) 
\nonumber
\\
&&~~~~~~~~ - 2\mu c_m\Big(\tilde{v}^a_3 + t^{ab}_2\omega_b+\frac{1}{3}(s_{2,1}-\frac{3}{2}\omega^2)\omega^a\Big)+c_m\Big(\Delta^{ab}E^c\nabla_c\omega_a
+2\omega^aE_ba^b-E^aa_b\omega^b\Big)~;
\nonumber
\\
&& \tau^{ab}_A = 4\mu c_m\Big(\tilde{t}^{ab}_3 - W^{c<ab>d}u_c\omega_d+2\omega^{<a}v^{b>}_{2,1}\Big)+2c_m\omega^{<a}v^{b>}_{2,2}+4c_m\omega^{<a}\Omega^{b>c}E_c
\nonumber
\\
&&~~~~~~~~~~+4c_mB^{<a}v^{b>}_{2,1}+2c_m\epsilon^{ijk<a}u_iE_j\Big(2t^{b>}_{2k}-R_{kl}\Delta^{b>l}\Big)-2c_m\nabla^{<a}F^{b>c}\omega_c
\nonumber
\\
&&~~~~~~~~~~+4c_ma^{<a}\epsilon^{b>ijk}u_iB_j\omega_k~.
\label{1.10}
\end{eqnarray}
The definitions of the several symbols are
\begin{eqnarray}
&& B^a = \frac{1}{2}\epsilon^{abcd}u_bF_{cd}; \,\,\,\ \omega^a=\epsilon^{abcd}u_b\nabla_cu_d; \,\,\,\ E_a = F_{ab}u^b; \,\,\ \Delta^{ab} = g^{ab}+u^au^b~;
\nonumber
\\
&& \sigma^{ab}=\Delta^{ac}\Delta^{bd}(\nabla_cu_d+\nabla_du_c)-\frac{1}{3}\Delta^{ab} \nabla_cu^c~,
\label{1.11}
\end{eqnarray}
and
\begin{eqnarray}
&& \Omega^{ab} = \frac{1}{2}\Delta^{ac}\Delta^{bd}(\nabla_cu_d-\nabla_du_c); \,\,\,\ t^{ab}_2 = W^{acdb}u_cu_d~;
\nonumber
\\
&& s_{2,1} = R+6R_{ab}u^au^b-6a^2; \,\,\,\ s_{2,2}=u_a\nabla_bF^{ab}~;
\nonumber
\\
&& v^a_{2,1} = \Delta^{ab}R_{bc}u^c-2\Omega^{ab}a_b; \,\,\,\ v^a_{2,2}=\Delta^{ab}\nabla^cF_{bc}~;
\nonumber
\\
&& \tilde{v}^a_3 = \epsilon^{abcd}u_b(\nabla_cR_{di})u^i+a^{(b}\Delta^{c)a}\nabla_b\omega_c+\omega^aa^2~;
\nonumber
\\
&& \tilde{t}^{ab}_3 = \Delta^{c<a}\epsilon^{b>dij}u_d(\nabla_iR_{jc}+2a_{i}t_{2jc}); \,\,\ \Phi_{ab}=R_{abcd}u^cu^d~,
\label{1.12} 
\end{eqnarray}
where we denote $V^{<ab>} = \Delta^{ac}\Delta^{bd}V_{(cd)} - \frac{1}{3}\Delta^{ab}\Delta_{ij}V^{ij}$ and $(ab)$ represents the symmetric combination. $a_a = u^b\nabla_bu_a$ is the acceleration perpendicular to $u^a$. Here $\zeta$ and $\eta$ are bulk and shear viscosities, respectively. These hideous expressions have been written in a much more compact and enlightening way in {\cite{Jensen:2013kka}} in terms of ``spin chemical potential''.  In this paper, I shall stick to the above forms.

\section{\label{relation}Vacuum state and the relation}
 This section will contain the main discussion of this paper. Here the role of Israel-Hartle-Hawking vacuum condition will be enlightened for deriving the algebraic relation (\ref{main}). For that first the null-null components of the stress-tensor will be found out from the expressions given in the preceding section and then I shall proceed towards the goal. 

  Note that the above expressions are very clumsy. So to proceed further let us consider a simple situation where the ($1+3$) dimensional background metric is static and spherically symmetric, in which case all the components of the stress-tensor take simple forms. In Schwarzschild like coordinates it is in the following form:
\begin{equation}
ds^2 =-f(r)dt^2+\frac{dr^2}{f(r)}+r^2(d\theta^2+\sin^2\theta d\phi^2)~.
\label{metric}
\end{equation}
The above metric has a timelike Killing vector whose vanishing of the norm gives the location of the Killing horizon. This is given by $f(r=r_0)=f(r_0)=0$. For our future purpose we define the null coordinates as $u=t-r_*$; $v=t+r_*$ where the tortoise coordinate $r_*$ is given by $dr_* = dr/f(r)$. The relation between the components of stress-tensor in the both coordinates are
\begin{eqnarray}
&& T_{uu} = \frac{f^2}{4}\Big(T^{tt}+\frac{2}{f}T^{tr}+\frac{1}{f^2}T^{rr}\Big)~;
\nonumber
\\
&& T_{vv} = \frac{f^2}{4}\Big(T^{tt}-\frac{2}{f}T^{tr}+\frac{1}{f^2}T^{rr}\Big)~; 
\nonumber
\\
&& T_{uv} = \frac{f^2}{4}\Big(T^{tt}-\frac{1}{f^2}T^{rr}\Big)~.
\label{1.13}
\end{eqnarray}
Next the components of $T^{ab}$, given by (\ref{1.06}) and (\ref{1.08}), will be evaluated for the metric (\ref{metric}). For that choose the comoving frame in which the components of velocity vector turn out to be
\begin{equation}
u^a=(\frac{1}{\sqrt{f}},0,0,0)~; \,\,\,\ u_a = (-\sqrt{f},0,0,0)~. 
\label{1.14}
\end{equation}
Then all the components of $\omega^a$ vanish. The non-vanishing component of the acceleration is $a_r = f'/(2f)$ and hence the norm is $a^2=f'^2/(4f)$ where the prime is the derivative with respect to $r$-coordinate. Using all these one can show that 
\begin{equation}
T^{tt}_{(0)} = \mathcal{E}(u^t)^2~; \,\,\ T^{tr}_{(0)}=u^tB^r \tilde{c}_{4d}T^2~; \,\,\ T^{rr}_{(0)}=f\mathcal{P}~,
\label{1.15} 
\end{equation}
and 
\begin{equation}
T^{tt}_{(A)} = \mathcal{E}_A(u^t)^2~; \,\,\ T^{tr}_{(A)}=u^tB^r\Big(-3c_A\mu^2-c_mf''+c_m\frac{f'^2}{2f}\Big)~; \,\,\,\ T^{rr}_{(A)}=f\mathcal{P}_A~.
\label{1.16} 
\end{equation}
Substituting the above results in (\ref{1.13}) and using $B^r=-u_tF_{\theta\phi}$, we obtain the required components of stress-tensor in null coordinates as
\begin{eqnarray}
&& T_{uu}=\frac{1}{4}\Big[f\Big(\mathcal{E}+\mathcal{P}+\mathcal{E}_A+\mathcal{P}_A\Big) + 2F_{\theta\phi}\Big(\tilde{c}_{4d}T_0^2-3fc_A\mu^2-c_mff''+c_m\frac{f'^2}{2}\Big)\Big]~;
\nonumber
\\
&& T_{vv}=\frac{1}{4}\Big[f\Big(\mathcal{E}+\mathcal{P}+\mathcal{E}_A+\mathcal{P}_A\Big) - 2F_{\theta\phi}\Big(\tilde{c}_{4d}T_0^2-3fc_A\mu^2-c_mff''+c_m\frac{f'^2}{2}\Big)\Big]~;
\nonumber
\\
&& T_{uv}=\frac{f}{4}\Big(\mathcal{E}-\mathcal{P}+\mathcal{E}_A-\mathcal{P}_A\Big)~,
\label{T}
\end{eqnarray}
where the relation between $T$ and equilibrium temperature $T_0$, given by $T=T_0/\sqrt{f}$, has been used.

    After obtaining the components, next step is to find the relation between the response parameter $\tilde{c}_{4d}$ and the anomaly coefficient $c_m$. This will be done by imposing the {\it Israel-Hartle-Hawking vacuum} condition. Before going into this discussion let me introduce the applicability of the three different quantum states \cite{Balbinot:1999vg} which are relevant for the metric (\ref{metric}). (i) Boulware vacuum: This vacuum is defined in such a way that both the {\it in} and {\it out} modes have positive frequency with respect to the Killing time in the Schwarzschild like coordinates. Therefore in the asymptotic limit $r\rightarrow \infty$, it is Minkowskian and hence the components of energy-momentum tensor in Schwarzschild like coordinates must vanish for this limit. On the other hand, near the horizon $T_{ab}$ is divergent. This state is usually used to describe the vacuum polarization around a static star whose radius is larger than that of the horizon. (ii) Unruh vacuum: Here the positive frequency {\it in} modes are chosen with respect to the Schwarzschild timelike Killing vector while the positive frequency {\it out} modes are defined with respect to the Kruskal $U$ coordinate. The Kruskal $U$ and $V$ coordinates are related to the null coordinates by the relations: $\kappa U = -\exp[-\kappa u]$ and $\kappa V = \exp[\kappa v]$, respectively where $\kappa = f'(r_0)/2$ is the surface gravity and so $T_{UU} = T_{uu}/(\kappa U)^2$ and $T_{VV} = T_{uu}/(\kappa V)^2$. This implies that the vacuum is Minkowskian in the $r\rightarrow \infty$ limit and so there is no ingoing flux; i.e. $T_{vv} = 0$. On the other hand $T_{UU}$ must be regular near the horizon which implies $T_{uu} = 0$ in the limit $r\rightarrow r_0$. This state is suitable for the evaporation of a black hole. (iii) Israel-Hartle-Hawking vacuum: In this state the {\it in} modes have positive frequency with respect to the Kruskal $V$ coordinate and {\it out} modes are positive frequency with respect to $U$. Therefore both $T_{VV}$ and $T_{UU}$ are regular near the horizon. Then $T_{vv}$ and $T_{uu}$ both must vanish in the limit $r\rightarrow r_0$. This state describes a black hole system is in thermal equilibrium  with the surroundings. In the present context, since the fluid is at thermal equilibrium with the black hole, the natural vacuum, one may think, is the Israel-Hartle-Hawking state. We shall show below that this will give the required relation.

As explained in the above, the definition of the Israel-Hartle-Hawking boundary condition leads to the vanishing of $T_{uu}$ and $T_{vv}$ near the Killing horizon \cite{Balbinot:1999vg}. Since in the near horizon limit (i.e. $r\rightarrow r_0$), the quantities $\mathcal{E},\mathcal{P},\mathcal{E}_A,\mathcal{P}_A, F_{\theta\phi},\mu$ all are finite, imposition of the condition $T_{uu}(r\rightarrow r_0)=0$ yields
\begin{equation}
\tilde{c}_{4d} T_0^2 = -c_m\frac{f'^2(r_0)}{2}~.
\label{c}  
\end{equation}
Now the equilibrium temperature, in this case, is given by $T_{0} = f'(r_0)/(4\pi)$ and hence the above reduces to (\ref{main}). Similarly, from $T_{vv}$ component we obtain the same relation. This was also obtained earlier in \cite{Jensen:2012kj,Jensen:2013rga} by studying the Euclidean partition function on a cone.

  Before concluding, let me  mention that in the above no use of the current has been done. For completeness, let us now find the components of the current in null coordinates and see how they behave under the Israel-Hartle-Hawking boundary condition on the current ($J_u\rightarrow 0$ and $J_v\rightarrow 0$ in the limit $r\rightarrow r_0$) or if one can extract more information about the fluids. The computation gives the form of the components as follows:
\begin{eqnarray}
&& J_u=-\frac{1}{2}\Big[\sqrt{f}\frac{\partial P}{\partial\mu}-6c_AF_{\theta\phi}\sqrt{f}\mu^2\Big]~;
\nonumber
\\
&& J_v=-\frac{1}{2}\Big[\sqrt{f}\frac{\partial P}{\partial\mu}+6c_AF_{\theta\phi}\sqrt{f}\mu^2\Big]~.
\label{current}
\end{eqnarray}
Note that they vanish near the horizon. So the vacuum condition is satisfied trivially and we do not have any new information from the expression for current. Also remember that the gravitational contributions in the diffeomorphism anomalies appear only in the even dimensional theory, like in $2n$ spacetime dimensions where $n=1,2,3,\dots$. For $2D$ case, we have a purely gravitational part whereas in $4D$ we have a mixture of both gravity and gauge contributions. It has been shown that the response parameter is related with the coefficient of the pure gravitational anomaly in $2D$ whereas, for the present case, it is connected to the mixed anomaly coefficient. Interestingly, in both cases, I have shown that these can be achieved by imposing the regularity condition of the stress-tensor in Kruskal coordinates near the horizon; i.e. by imposing the Israel-Hartle-Hawking vacuum. Moreover the relation is quite general in structure. Since in anomaly equations the pure gravitational part always appears in $2j$ with $j=1,3,5,\dots$ spacetime dimensions whereas mixed anomaly comes in $2k$ with $k=2,4,6,\dots$ dimensions, we expect the similar connection will happen in other even dimensional theories like in $2D$ and $4D$ when one will use the same regularity condition. More explicitly it may be noted that in $2D$ and $4D$ theories, the anomaly polynomials, which lead to the respective anomaly equations, contain the first Pontryagin class \cite{Jensen:2012kj,Jensen:2013kka} and the response parameters are related to the coefficients of this function. Furthermore, it has been already shown in \cite{Jensen:2013kka} that the anomaly polynomials in higher even dimensions are given by the higher Pontryagin class. Therefore, in these higher dimensional cases the response parameter will again be related to the coefficients of Pontryagin class. The explicit expressions, of course, will be obtained when one will write the stress-tensor and the current and use the regularity conditions on them near the horizon. This I leave for future work.

\section{\label{conclusions}Conclusions}
  In this brief report, the earlier analysis \cite{Majhi:2014eta} for ($1+1$) dimensional anomalous fluid has been extended to ($1+3$) dimensions case. The theory for the present case has both the gravitational as well as $U(1)$ gauge contributions which are mixed in the anomaly equations. I showed that the Israel-Hartle-Hawking vacuum is the relevant vacuum condition in the context of anomalous hydrodynamics. Use of it on the constitutive relations gave us the correct anomalous contribution to the response parameter in the first order derivative term.

    It is worth to mention that the other two vacua (Boulware and Unruh) which are asymptotically Minkowski in the $r\rightarrow\infty$ limit, do not describe a fluid at thermal equilibrium with the black hole and hence they are not suitable to discuss the present situation. Whereas, the Israel-Hartle-Hawking is a natural choice to describe this equilibrium system. Interestingly, this gives the correct relation as obtained earlier by using the Euclidean partition function approach \cite{Jensen:2012kj,Jensen:2013kka}.

  One may note that the approach works well in both two and four dimensional theories. The analysis gives a strong evidence that the Israel-Hartle-Hawking vacuum plays a significant role in anomalous fluid dynamics. Also, due to the simplicity of the method, one can apply to any arbitrary dimensional theory. Incidentally, it has been argued in my earlier work \cite{Majhi:2014eta} for two dimensional case that such an approach is quite similar to derivation of Cardy formula. In other words the Cardy formula played a role in the fixing of the relations in the derivative expansion approach as well in the vacuum approach. That is an important point to justify the reason that the results from these approaches agree, having a common origin as the Cardy formula. Hope in future, I will be able to explore more on this topic and give a deep understating over the importance of vacuum states. The investigations in these directions are in progress.

\vskip 9mm
\noindent
{\bf{Acknowledgements}}\\
\noindent
I thank Rabin Banerjee, Shirsendu Dey and Kristan Jensen for several useful comments and suggestions. The research of the author is supported by a Lady Davis Fellowship at Hebrew University, by the I-CORE Program of the Planning and Budgeting Committee and the Israel Science Foundation (Grant No. 1937/12), as well as by the Israel Science Foundation personal Grant No. 24/12.

\end{document}